# A SEMIEMPIRICAL DYNAMICAL MODEL TO FORECAST THE PROPAGATION OF EPIDEMICS: THE CASE OF THE SARS-COV-2 IN SPAIN

Juan C. Mora[1], Sandra Pérez[2], Ignacio Rodríguez[3], Asunción Núñez[3], & Alla Dvorzhak[1].

[1] Dept. of Environment. CIEMAT. Avda. Complutense, 40. 28040. Madrid. Spain.

[2] Sercomex Pharma. C/ Pollensa, 2. 28232. Las Rozas de Madrid. Spain.

[3] Dragonfly Innovation Technologies. 145-157 St. John Street, London, EC1V 4PW. UK.

20$^{th}$ of April, 2020

## ABSTRACT

A semiempirical model, based in the logistic map, has been succesfully applied to forecast important quantities along the several phases of the outbreak of the covid-19 for different countries. This paper shows how the model was calibrated and applied to perform predictions of people needing to be hospitalized, needs of ventilators, or the number of deaths which would be produced. It is shown specifically the results obtained in the case of Spain, showing a prediction of diagnosed infected and deaths which will be observed after the ease of the total lockdown produced the 13$^{th}$ of March. Is also shown how this model can provide an insight of what the level of infection in the different regions of Spain is forecasted.

The model predicts for Spain for the end of May more than 400,000 diagnosed infected cases, number which will be probably higher due to the change in the possibilities of performing massive number of tests to the general population. The number of forecasted deaths for that date is 46,000± 15,000.

The model also predicts the level of infection at the different Spanish regions, providing a counterintuitive result in the cases of Madrid and Catalonia as the result shows a higher the level of infection at Catalonia than the level at Madrid, according with this model.

All of these results can be used to guide policy makers in order to optimize resources and to avoid future outbreaks of the covid-19.

**INTRODUCTION**

A new respiratory disease, initially dominated by pneumonia, and caused by a coronavirus, was detected in China at the end of 2019. It was initially named by the World Health Organization (WHO) as 2019-nCoV (Zhu et al., 2020) and renamed in February 2020 by the International Committee on Virus Taxonomy as Severe Acute Respiratory Syndrome coronavirus 2 (SARS-CoV-2) (Gorbalenya et al., 2020), recognizing it as a sister of the SARS-CoV viruses. The same day the WHO named the disease as Coronavirus Disease 2019 (COVID-19) (WHO, 2019).

Many efforts have been done since then to appropriately model the spread of the disease in the world and in the different countries where infection arrived. Modelling the epidemics have many practical uses: preparation of national health systems; buying the necessary sanitary material; forecasting when the saturation of the health system could occur; when and to what extent apply Non Pharmaceutical Interventions (NPI) (Feng et al., 2010); or the time when those countermeasures can be relaxed. These theoretical approaches to predict the evolution of epidemics often use compartments models as simple as SIR (Susceptible, Infected, Recovered – sometimes called Removed) (Kermack and McKendrick, 1927), but in many occasions complexity is increased to include different phenomena. For example including the individuals who can infect others without presenting symptoms, as done in the SEIR models (Susceptible, Exposed, Infected, Removed); or assuming that recovered are not immune and can be infected again, as in the SEIRS models (Susceptible, Exposed, Infected, Removed, Susceptible); or including the deaths and births during the epidemic for long term infections, as those due to influenza; or often including compartments to distinguish deaths, recovered, hospitalized, and other situations by means of ratios (see for instance Brauer, 2008 for further information).

Since the SARS-CoV-2 outbreak many efforts have been carried out to adapt these compartment models to the real behavior of this particular virus. For example, a conceptual representation of covid-19 disease propagation, performed by the authors, can be seen in figure 1, where immunization of recovered is assumed only for a given period (12 months is typical for the immunization against the coronavirus causing common cold).

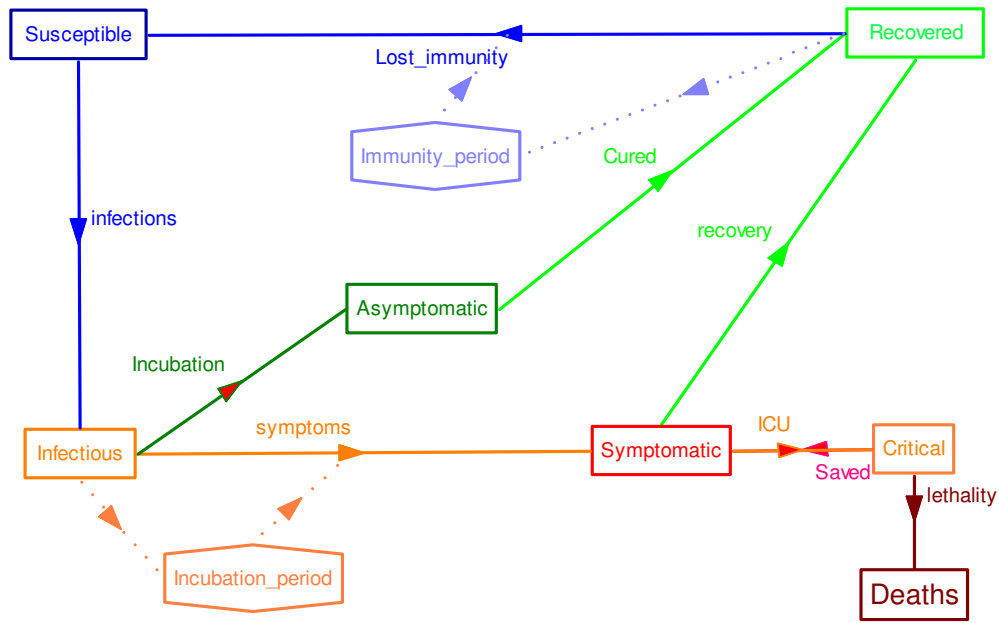

Figure 1.- Example of a SIR type compartment model adapted to simulate the behavior of SARS-CoV-2 in humans.

Due to the difficulty of adapting the parameters of these models at early stages of the epidemic spread, many times those efforts lead to wrong conclusions, for instance predicting the timing of the epidemic[1], or including huge uncertainties which are not well understood by the public[2]. Sometimes the authors of such predictions blame the quality of the statistical data. However, this quality is severely affected by the urgency of the epidemic and could not probably be avoided in this or in any future outbreaks of epidemics. The continuous publications of medical and epidemiological studies on the covid-19, and the associated virus, don't make it easy to extract good quality information to adapt the models. But it must be accepted that this situation will be always the case – or even worst - when new diseases appear.

---

[1] See the reports for the Spanish evolution published at https://covid19.webs.upv.es/informes.html (consulted on April 15th 2020 - in Spanish), where the peak of cases were predicted to happen at the end of May. In reality the peak was given on the 2$^{nd}$ of April 2020 in Spain.

[2] For instance the predictions published at https://covid19.healthdata.org/ (consulted on April 15$^{th}$ 2020) provided uncertainties for the results of one order of magnitude for the beds needed at USA for the 14$^{th}$ of April.

One clear example of problems associated with future predictions of the covid-19 behaviour is the uncertainty about the influence of ambient temperature or humidity, which would influence the seasonality of the disease (Wang et al., 2020). In the early stages many aspects of the behavior of the SARS-CoV-2 virus were associated with previous studies on similar viruses as the SARS virus, as could be the resilience in fomites (Kampf et al., 2020) or the immunization of patients recovered from it (Prompetchara et al., 2020). While writing this paper many aspects are under investigation, but in this line it is believed that, as happens with other human coronaviruses causing diseases like the 15% of the common cold cases (Pelczar, 2010), immunity will remain for a brief period, of the order of months.

During the outbreak of the epidemic in Spain several models were checked and a follow-up of the many results published were performed to advice Spanish national authorities (Mora, 2020). The best results were obtained by using a semiempirical approach presented herein, which has the advantage of doing accurate predictions with a minimum amount of information available during this epidemic, which possibly will be the situation in future outbreaks.

This paper presents the mathematical development and results obtained using this semiempirical model, focusing into the Spanish case, while presenting also some results obtained in other countries.

**MATERIALS AND METHODS**

The semiempirical forecasting model here presented was very accurate at the early stages of the epidemic, and with some calibration, was also able to do accurate predictions after NPI were applied in the countries, specifically total lockdown. Although this model does not make use of the basic reproduction number R0 used in the SIR, it was derived from the data and predictions made in 10 different countries, ranging from 1.3 to 4.2, which is in accordance with previous estimations (Liu et al., 2020).

This model basically applies the well-known logistic map which has been applied to many examples of growth of populations and widely studied due to its chaotic behavior dependent of the applied single constant (figure 2 shows a typical fractal created with the logistic map as a function of r). Any r in the equation below 3 will provide a typical behavior and reach equilibrium in the long term.

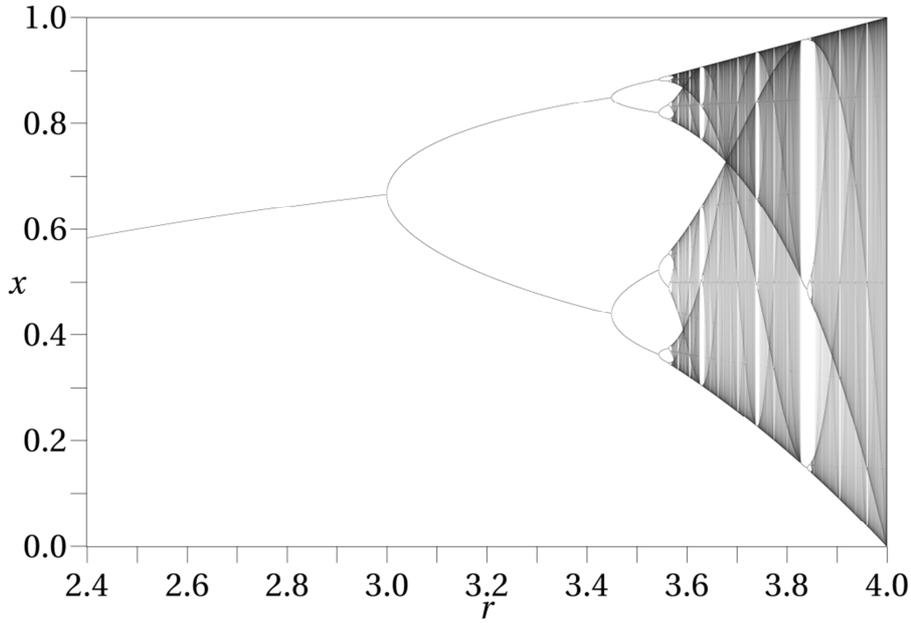

Figure 2.- Bifurcation diagram for the logistic map as a function of r.

The basic equation used to determine the number of infected diagnosed cases, based in the logistic map, is given by equation 1.

$$I(t) = r \cdot \left[1 - \frac{I(t-1)}{N}\right] \cdot I(t-1) \qquad \text{Equation 1}$$

Being I(t) the number of infected diagnosed cases at day t, I(t-1) the infected diagnosed cases of the precedent day t-1, r is the ratio of the logistic map (named hereafter daily infection rate), and N the number of individuals susceptible to be infected. Take into account that the number of susceptible individuals used here shouldn't be the same used for modellers applying SIR type models. The sub-index n will be used below to indicate the n-th day after the outbreak.

The behaviour unaltered of this function gives rise to the logistic function and the typical sigmoid shape (see figure 3).

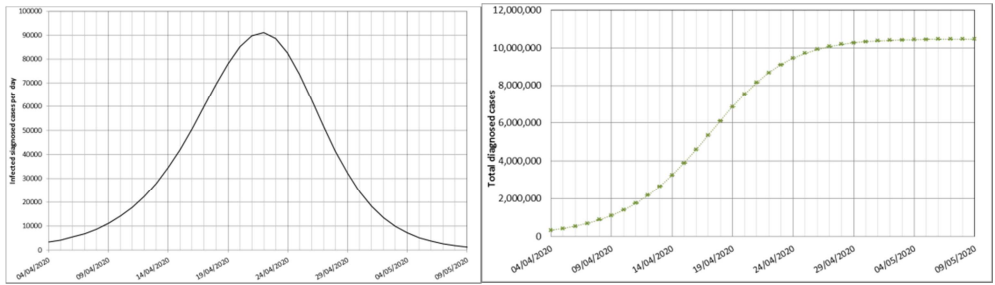

Figure 3.- Example of infected diagnosed cases daily rate (left) and its cumulative function of the total cumulated cases (right).

Other quantities needed to provide practical forecasting were the number of those infected cases who would need to receive medical attention at the hospital ($H_n$), the number of those who would need intensive care ($C_n$) and the number of deaths ($D_n$) at times t. $H_n$ and $D_n$ are calculated as a fraction of the number of the diagnosed infected cases at time t ($I_n$), and $C_n$ as fraction of $H_n$. Obviously the number of recovered patients ($R_n$) is given by the fraction (1- $D_n$). The fraction used to calculate $D_n$ in this way is often mentioned as case fatality rate (CFR), determined as $CFR = \frac{D_n}{D_n + R_n}$. This was proved to be more practical during the outbreak than other approaches as the mortality rate defined for the whole population which can be only experimentally determined at the end of the epidemic. A delay from the number of diagnosed cases to the deaths must be introduced to represent the time since the death to the diagnoses to the report of the death to the authorities.

These numbers are crucial to policy makers in order to take well founded decisions.

*Initial parametrization*

Unlike other viruses, SARS-CoV-2 is assumed to be able to infect with the same probability every human, disregarding sex or age. Moreover, no immunization is previously acquired, by natural or artificial means. Therefore, N was initially assumed to be the whole population of a given region, whatever the size of the region is. For the shake of simplicity, and for the examples which will be shown thereafter, the total population of a country (or of a region, as the autonomous communities in Spain) is used for N.

All the other parameters are determined empirically by averaging the available information in a studied country, even if the available data cover only few days, or by taking the information from previously affected countries.

The daily infection rate r can be dynamically determined taking all the data collected or by using a given period to average the data, using the number of diagnosed infected cases of day n+1 ($I_{n+1}$) divided by the same number at day n ($I_n$). For instance, averaging the daily infection rate during the 15 days from the 20$^{th}$ of February to the 5$^{th}$ of March, r = 1.35 was obtained for Spain and r = 2.09 for Italy. For the same period of 15 days taken at the beginning of the outbreak r = 1.55 was obtained for France, 1.34 for the USA, 1.35 for Germany, 1.67 for South Korea or 1.32 for the UK.

The fraction of the diagnosed infected who need to be hospitalized ($H = \frac{H_n}{I_n}$) was dynamically determined from all the previous data acquired at each region (or state, or country) and averaged for the whole period since the beginning of the epidemic. The same was done for the number of patients needing an ICU ($C = \frac{C_n}{H_n}$). An initial factor of patients needing an ICU from the number of infected cases H·C = 0.05 to 0.15 was computed from the studies at Eastern countries.

As the reported data for the diagnosed infected cases are usually given as cumulated since the beginning of the epidemic, obviously all the other quantities will be obtained also as cumulated functions. This was demonstrated to be a problem when different regions decided to report different quantities and then it made it impossible to obtain real cumulated data. For some cases it is interesting to obtain the daily rates, for instance to calculate the day where the maximum (peak) of infections or deaths would occur.

For the CFR, the value achieved in the equilibrium in China was used (of a 0.0578 as measured the 4$^{th}$ of March - see figure 4 to see the evolution of the CFR in China). The big rates observed at the beginning of the epidemic can be due to the high lethality on the more vulnerable population or due to the lack of knowledge on which medical treatments should be applied, what improves with time. A similar evolution was observed in other countries, although with a slowest decrease in the cases of European countries (at the time writing this paper CFR for USA was 0.3998, for Italy was 0.3557, and for Spain was 0.2052). The time delay from diagnosed cases to deaths must be calibrated, as in the initial stages the observed delay was higher for all the countries and drastically reduced after some days after the outbreak. An initial delay of 5 to 10 days was often observed, which, as explained, includes the time needed to perform the needed tests and register the case as a positive death caused by the covid-19.

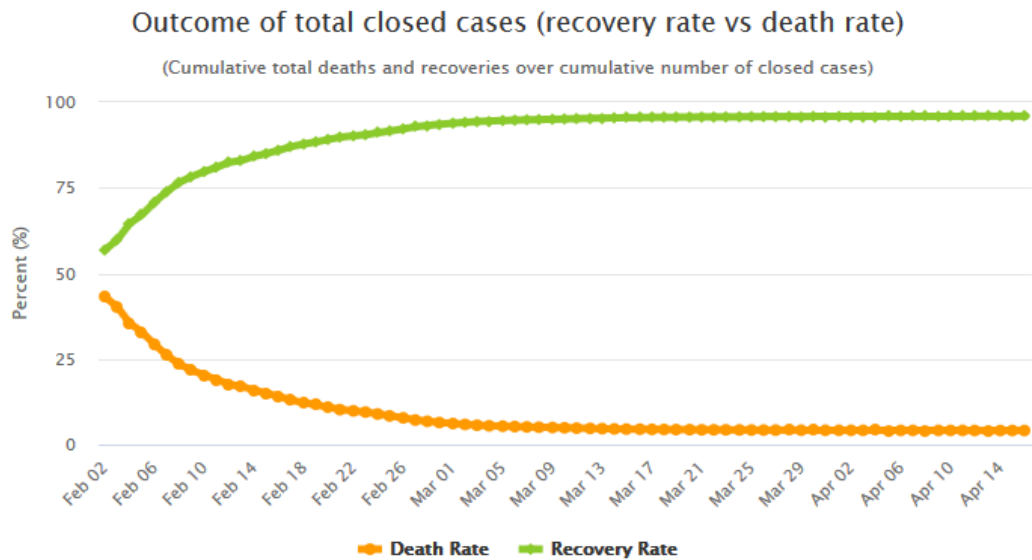

Figure 4.- Evolution of the CFR (orange) and recovery rate (green) at China[3].

Examples of the results obtained in the initial stages of the outbreaks, before the application of NPI countermeasures, are presented below.

*Parametrization after lockdown*

The more used NPI since the outbreak at China has been the lockdown, partial in some countries and total in many others, including the region of Hubei in China (58.5 million inhabitants), Spain (46.6 million inhabitants) or Italy (60.4 million inhabitants).

In the countries the total population was used initially for N, but after the lockdown, the number of people who were already infected, or were still in contact with people who can infect them, is fixed and should be smaller than the initial N. This number cannot be determined before the lockdown and should be calibrated with the observed cases of diagnosed infected. However, a first estimation which provided good results was considering the number of diagnosed infected cases, estimated with the model, 14 days after the lockdown (14 days was assumed to be the incubation period for the covid19) and multiplying that estimation by a factor of 10, which would provide the total infected (symptomatic or asymptomatic). This method provides a rough estimation which needs further refinements when new data is obtained, however it provided valid estimations for forecasting the time when the maximum (peak) for daily new cases of diagnosed cases or deaths could be expected.

---

[3] Figure taken from https://www.worldometers.info/coronavirus/country/china/ which demonstrated a continued and confident series of data along the time (consulted the April 17th).

The daily infection rate was observed to decrease, from the rate before the NPI was applied (typically around 1.3) to nearly 1.0, as observed at South Korea. The same behaviour was observed in every country and at every scale. So the daily infection rate r, after lockdown, can be well fitted by means squares to the curve given in equation 2, either for a region or a country.

$$r = 1 + A \cdot e^{-\alpha \cdot t}$$
Equation 2

Where t is the time in days since the lockdown and A and α are constants determined empirically at every location. Table 1 shows some of the values for A and α for different countries and examples of smaller scale regions within Spain (Andalucia and Catalonia were chosen for this example because they are the two more populated regions in Spain).

Table 1.- Values obtained by fitting equation 2 to the experimental data in different regions or countries. *Data used for the fitting from worldometer[4]. **Data used for the fitting from the Spanish official source of information[5].

| Region | A | α |
|---|---|---|
| South Korea* | 0.226 | 0.235 |
| Italy* | 0.293 | 0.070 |
| UK* | 0.326 | 0.051 |
| Spain** | 0.295 | 0.074 |
| Andalucia** | 0.366 | 0.096 |
| Catalonia** | 0.491 | 0.109 |

---

[4] Source of information https://www.worldometers.info/coronavirus/ (consulted on April 17th)
[5] Source of information Instituto de Salud Carlos III (ISCIII): https://covid19.isciii.es/.

The number of individuals susceptible to have been infected before the lockdown (N), and the constants A and α cannot be known prior to the lockdown, as different societies behave differently under the same instructions and also different governments provided slightly different instructions. So the only chance to obtain good predictions after the lockdown is to wait for several days to obtain experimental data to be used to fit the curve under the equation 2. It should be also pointed out that, as sometimes different sources of information provided data shifted in time or just different, the fitting could provide different values for the parameters.

For the parameters H and C empirical averaged values from the studied region were used. Also for the CFR empirical averaged values were used, as the initial value taken from China was well overpassed in many European countries, although with the time it is expected to tend to the same equilibrium value observed at China. In Spain for instance, the values obtained the $6^{th}$ of April were H = 0.467 and C = 0.1497, what indicates that an elevated rate of diagnosed cases needed to be hospitalized, or more probably that only severe cases were diagnosed at the hospitals needing about half of them to be hospitalized. Also a high percentage of the patients at the hospital (almost a 15%) needed at some time to receive intensive care using ventilators, what was in agreement with the observed pattern at China and other countries. In this case H·C = 0.07 (7%) which is within the initial range observed for the patients needing an ICU. As all parameters are dynamically calculated every day, the predictions can be slightly altered by the new data provided by the sources of information.

Finally, the CFR was also fitted to the real reported data for the region under study. The experimental CFR obtained in different countries was well above the equilibrium value from China. In the UK, Belgium, Spain or Italy a CFR of nearly 0.12 was obtained at the time this paper was written. The delay applied from the diagnosed infected cases to the number of deaths was reduced to 3 days, giving good results in general.

*Parametrization after easing lockdown restrictions*

When a region or a country decides to relax the confinement countermeasure, the parameters need a new calibration to take into account the new situation.

Two are the driving parameters which should be recalibrated: N an r. If the lockdown is completely abandoned, then N will return to be again the whole population of the region. However, this was not the situation in every country. In Spain for instance a large part of the population remained confined after the 13$^{th}$ of April where some relaxation was adopted. Therefore, the parameter can be only inferred after some data are collected. For r the ideal would be to have a value as close to 1 as possible before easing the countermeasures, as was the case in South Korea since the 12$^{th}$ of March. However a value from 1.01 to 1.03 was observed in most of the countries where also good practices were taken, as Germany or Norway in Europe.

These values can be used to perform conservative assessments on how the situation would evolve after easing the lockdown in the countries or to design the best strategy to achieve reasonable levels of new cases of diagnosed infected, hospitalized or patients needing an ICU in order not to saturate the health system of a country.

## RESULTS

As pointed out before, in general 3 phases have been observed in the different countries:

1.- An initial phase where no severe restrictions were applied.

2.- A second phase where severe NPI were applied.

3.- A last phase where relaxation of the NPI is assumed.

Some examples are here presented of the application of the model with the parametrization presented above to show the performance of the model, as a comprehensive compilation of all the countries and regions worldwide cannot be presented here.

*Initial phase*

As explained, the initial phase is considered before any NPI countermeasure is applied. The example chosen for this case was Italy. Figure 5 shows the number of total cases of infected diagnosed and the total number of deaths up to the 10$^{th}$ of April, where the total lockdown was established in the nation.

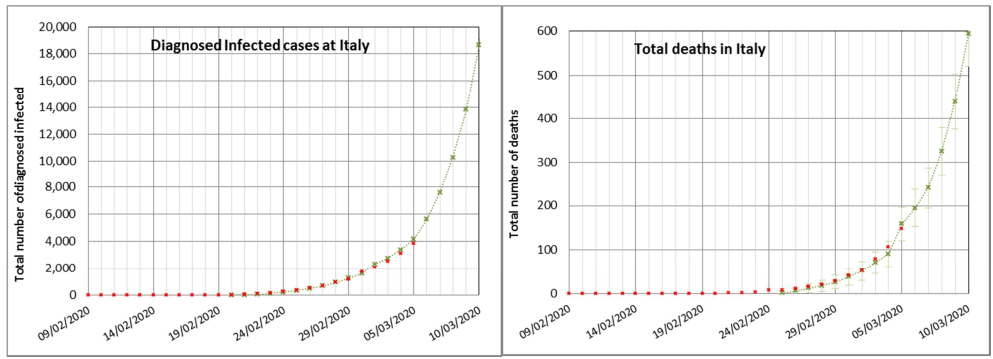

Figure 5.- Total diagnosed infected cases and total deaths in Italy from the 9th of February to the 10th of March.

During the initial stage the only parameter which was dynamically calibrated, to follow the real data, was the delay from the number of infected to the number of deaths, as was explained before, up to 2 days' delay applied the 5th of March. In this specific case, the forecast indicated a number of infected cases of over 18,000 and a number of deaths of nearly 600 to occur 5 days later. The real number of diagnosed cases was 10,149 (77% difference), however the real number of deaths was 631 (5% difference). Although the difference in diagnosed infected cases is big, it must be pointed out that several factors affected this number, as the under-prediction of positive cases, which was demonstrated for many countries. However, the accuracy for the number of deaths was very good as was the prediction on the number of patients needing an ICU (900-2700).

Very similar results were found for other countries as Spain, the UK or the USA.

These results were good to have an early idea of when the governments would apply extreme NPI like the lockdown, to predict final consequences in the case of no NPI applied and to prepare for the need of ventilators.

*NPI (total lockdown) phase*

As stated, for this second phase all factors were recalibrated, including a fitting of the daily infection rate to the curve given in equation 2. For the number of susceptible individuals N an initial estimation was carried out using the results of the model 14 days after the lockdown. As example the case of Spain is presented in figure 6.

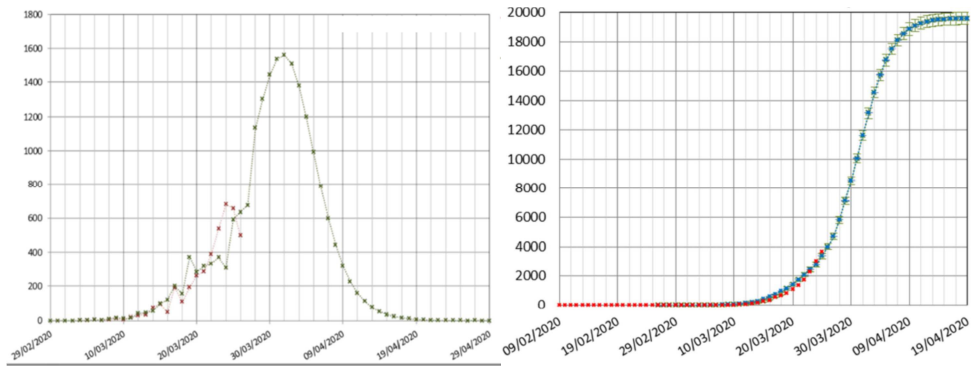

Figure 6.- Estimation of the daily rate of deaths (left) and the total number of deaths (right) predicted the 26$^{th}$ of March in Spain (Mora b, 2020). Real data in red, predictions in green or blue.

As shown, the total number of deaths by the end of April was predicted to be of about 20,000, while the peak for the daily deaths rate was predicted to occur for the 2$^{nd}$ of April. Although this paper is being written before the end of April, the number of deaths has already reached that number. The maximum (peak) of the daily rate of deaths was in fact achieved between the 1$^{st}$ and 2$^{nd}$ of April. In that case it was also predicted the need of 6,700 ventilators and ICU beds by the 30$^{th}$ of March, while the real number achieved that day was of 5,667 (18% difference). The prediction in that case was slightly conservative.

*Easing of NPI (unlocking) phase*

The last phase is actually occurring in many western countries in this moment, and happened days before in eastern countries as the Hubei region in China. In this case, only predictions based in some hypothesis can be presented in this study, as the real effect will be observed in some weeks from now. Again the case of Spain is presented here as example (figures 7, 8 and 9).

The day 13$^{th}$ of April a partial ease of the total lockdown was applied in Spain. As this paper is written the 17$^{th}$ of April only data up to the 16$^{th}$ of April are available and therefore there are not enough data to perform a means squares fitting. So only some reasonable hypotheses can be applied to determine N and r. As the total workforce in Spain is around 20 million people, and only some industries were allowed to begin again their activities, it was assumed that about a 20% of the total workforce (4 million workers) went back to work. As those people could also infect their families, assumed to be of 2 more members as average, a total N was taken of 12 million. The daily infection rate r was set constant as a 3%, which was the rate at the time of the unlocking.

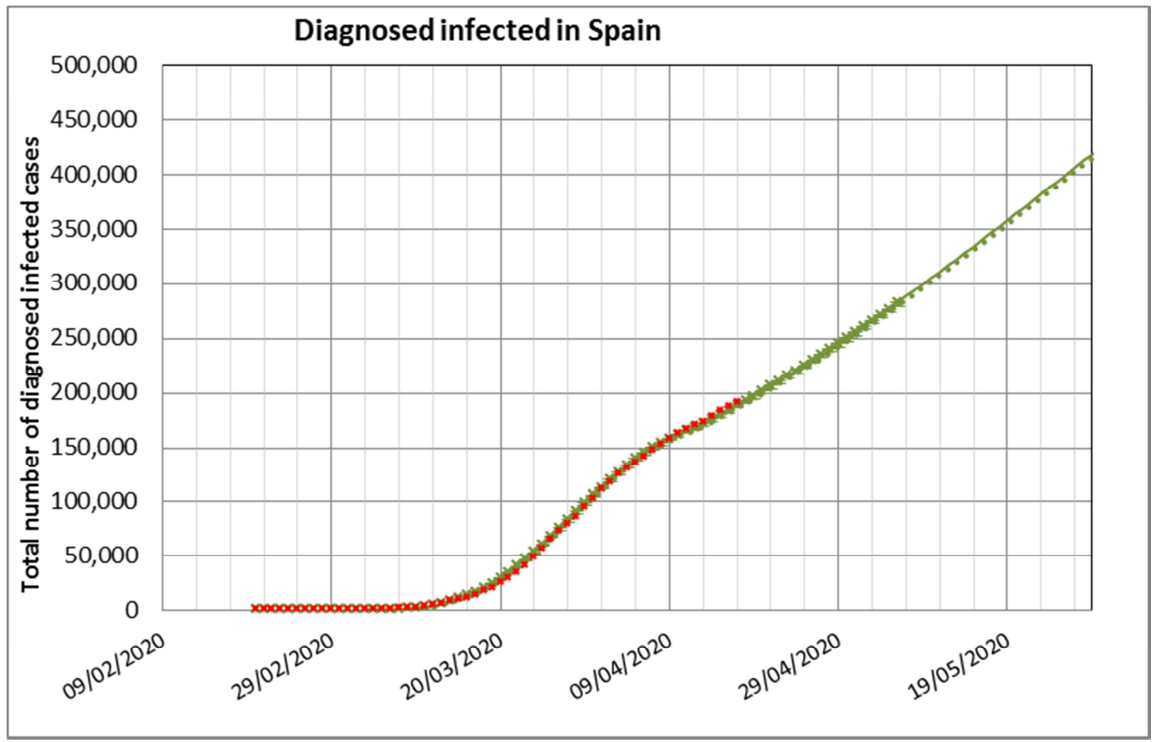

Figure 7.- Total number of diagnosed patients of covid-19 in Spain. Real data in red, predictions in green.

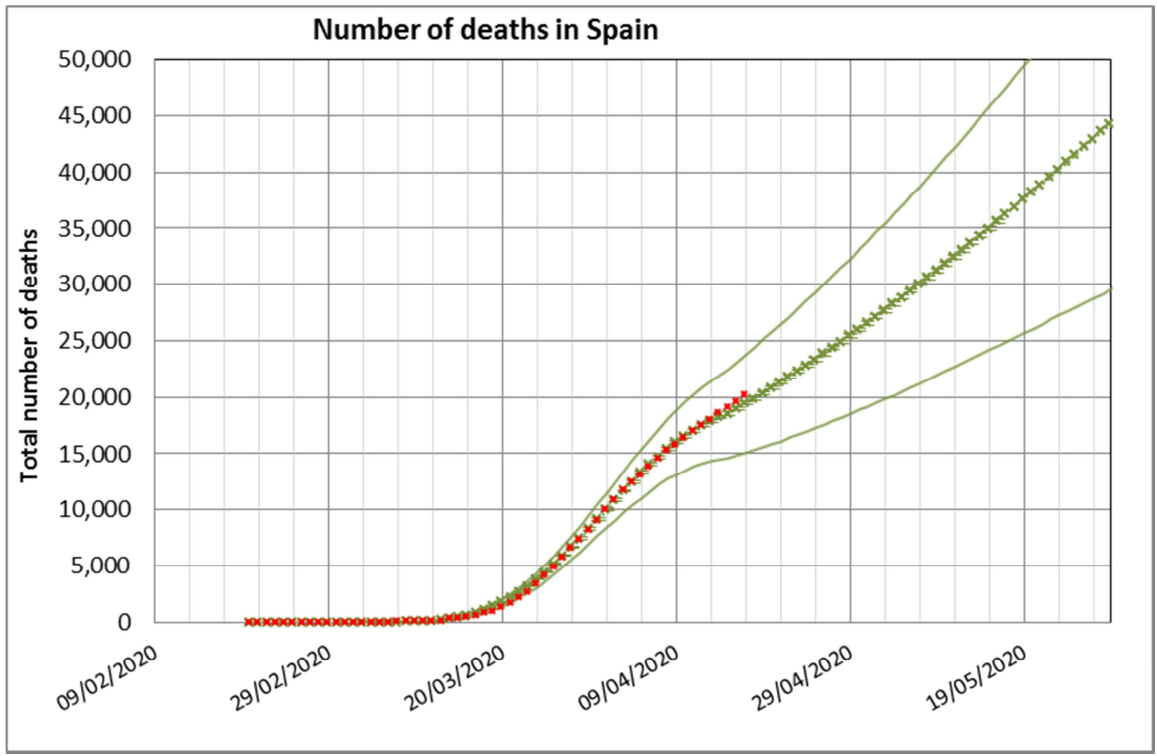

Figure 8.- Total number of deaths in Spain. In red the official data, in green the forecasted data based in the official previous reports together with the associated band of uncertainty (CL 95%).

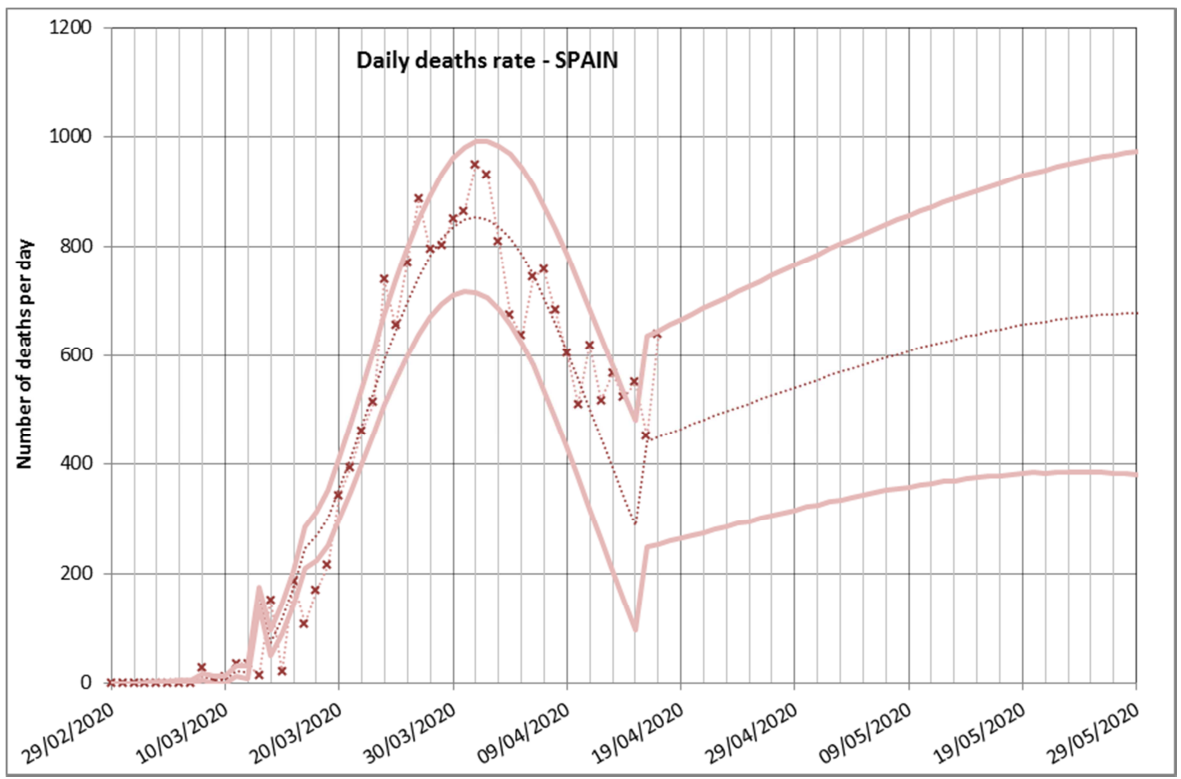

Figure 9.- Daily deaths rate in Spain. Red dots are real data; red lines are the forecasted daily deaths with the associated bands of uncertainty (CL 95%).

According with these predictions, some important consequences in easing the locking of a big number of the confined population are extracted.

First of all, the number of diagnosed infected people will increase with time continuously during the rest of the month of April and the whole month of May. In fact, if there are no changes in the ease of the confinement or in the statistical methodologies of counting infected or deaths, there will be a peak in the daily rate of infected by the 28$^{th}$ of May and a peak in the daily rate of deaths by the 1$^{st}$ of June. The total number of deaths in Spain by that date will reach the 46,000 ± 15,000.

Of course, these predictions do not consider important changes, as a sudden increase in the temperatures which would hopefully reduce the infectivity of the SARS-CoV-2, or the discovery of a vaccine - which seems extremely difficult in a so reduced period of time. During this time possibly the treatment of hospitalized patients will improve and therefore a smaller fraction of them would need intensive care with ventilators.

And, as mentioned, all of these predictions lose their validity if there are changes in the situation, as additional people leaving the confinement, relaxation of the use of personal countermeasures (gloves, masks), etc.

*Percentage of infected population in the regions*

There is an additional result which can be extracted from the use of this model to forecast the evolution of the pandemic. The need of calibrating the model by fitting the parameter N, which is the number of people susceptible to be infected, offers the possibility of using that number to infer the percentage of the population in a country or region which could have been infected, most of them showing no symptoms.

As an example these numbers were extracted for the autonomous communities (administrative regions) in Spain and transformed to three levels of infection (low – below 15% -, medium – from 15% to 30% and high – above 30%), giving rise to the result shown in figure 10.

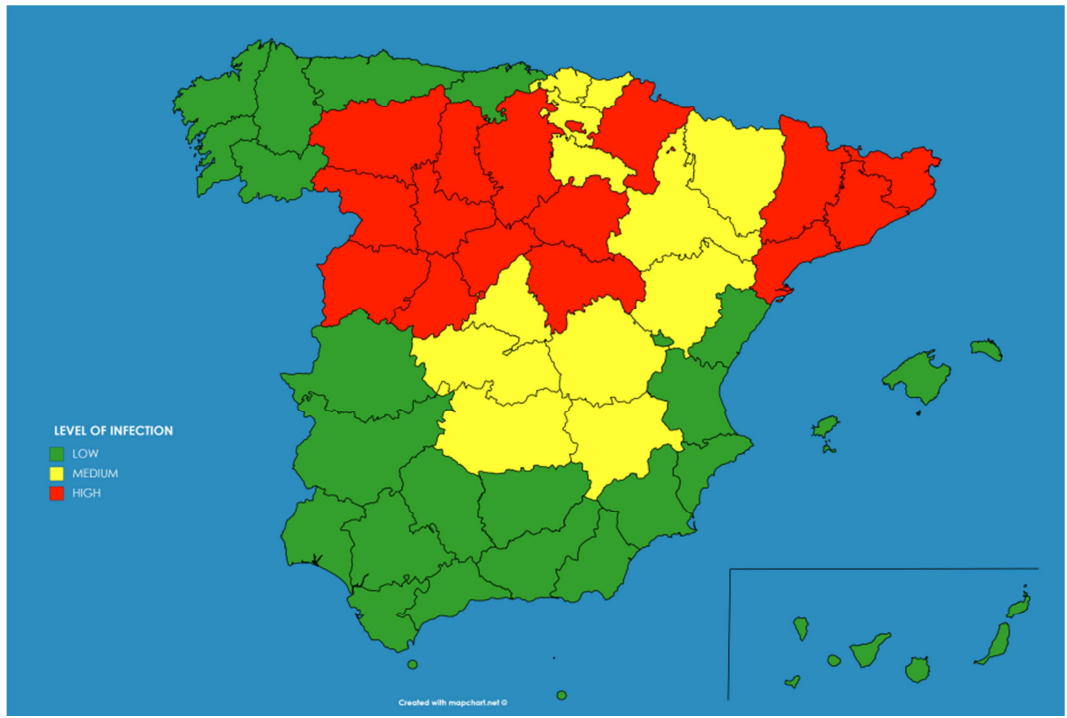

Figure 10.- Map of Spain showing the level of infection at each region forecasted with the model described in this paper.

This result has an important use for the authorities, as the population which have a high level of infection by the SARS-CoV-2 have probably developed immunity against the virus (at least temporarily as discussed before), and therefore there is no chance for them to be infected again in the future. And on the contrary, there is a bigger chance of being infected in those regions where the percentage of the infected population is smaller.

The model provides counterintuitive results in this case. As observed, the capital, Madrid, presents a medium level of infection of the population, whereas Catalonia shows a high level. In fact both, the number of diagnosed cases and deaths in Madrid was higher at the time of writing this paper, what would imply a higher level of infection in Madrid. However this could be explained in the different criteria applied in different regions. For example Catalonia decided to include in the statistics the deaths of people occurred in their homes and not only in the hospitals, while it was unknown if Madrid was including those deaths already in the statistics. The same occurs with the number of diagnosed infected cases, as there is observed an increase in the number of tests performed on people who finally doesn't need a hospital in Madrid (more than 84% don't need hospitals), the percentage in Catalonia of diagnosed population finally needing a hospital was still around a 65% (only a 35% didn't need a hospital).

For most of the other regions the numbers were in a good agreement with the expectations.

In any case, these predictions of the model can only be validated by carrying out an extensive program of tests to the population to confirm the percentage which was already infected.

**DISCUSSION AND CONCLUSIONS**

A methodology based in the application of the logistic map in different phases of the covid-19 epidemic was applied. This methodology provided good results in forecasting the evolution of the disease in every country where it was applied.

As still there is a long way to discover a vaccine to the SARS-CoV-2 or a cure of the associated disease covid-19, the only way of reducing the consequences of the epidemic is the optimum application of NPI.

The use of non-pharmaceutical countermeasures, as the total lockdown, showed their effectiveness during the period they were applied. However, easing the countermeasures will allow new outbreaks of the infection to appear. This situation forces the need of applying many simultaneous techniques to reduce the effect of the disease. One of those could be the application of the methodology described in this paper to provide early alert of those outbreaks in small units of population, allowing an optimization of sanitary resources and reducing the effect of future NPI if applied locally.

The aim of any policy of applying the NPI, or after easing their application, should be looking for daily infection rates of the range of 1.01 to 1.02 (1% to 2%), which, as shown in countries like South Korea, would provide acceptable levels of people needing care in hospitals or dying. And the more important, it makes almost impossible to reach to saturation levels of national healthcare systems.

A forecast for the next month on the number of diagnosed infected cases and deaths in Spain was presented, providing a number of deaths in Spain of 46,000 ± 15,000. These numbers, and also the number of diagnosed infected, can be altered due to new statistical methods, as including deaths at home, or the performance of a bigger number of tests in the non-symptomatic population.

Finally a prediction of the percentage of the population infected in the different regions of Spain was performed by using the suggested semiempirical model. These predictions have been shown in a map of Spain showing those regions where the infection level was low, medium or high. Some results obtained with this methodology were not intuitive according with the official information. The more counterintuitive result probably being the higher level of infection of Catalonia compared with Madrid region. This result can be only validated by the use of massive testing in the whole Spanish territory.

As this epidemic is still ongoing many situations would alter the predictions here presented. The discovery of a vaccine for the virus or a cure of the associated disease will deeply change the numbers. Changes in the weather will demonstrate the seasonality of the disease, but then new outbreaks will be expected for the season after the summer. A continuous watch of the disease is needed to provide adequate forecasting which can be used by policy makers.

## FUNDING

This work has not received any funding.